\documentclass[12pt,a4paper]{article}

\usepackage{amssymb}
\usepackage{amsmath}
\usepackage{amsfonts}
\newcommand{\imag}{\mathrm{i}\,}

\newtheorem{prop}{Proposition}
\title{Classification of quantum superintegrable systems with quadratic integrals on two dimensional manifolds}
\author{C. Daskaloyannis\thanks{e:mail address: daskalo@math.auth.gr}
and Y. Tanoudes\\
{Mathematics Department}
\\
Aristotle University of Thessaloniki\\
54124  Thessaloniki- Greece
}

\date{July 2006}

\begin{document}

 \maketitle

\begin{abstract}
There are two classes of  quantum integrable systems  on a
ma\-ni\-fold with quadratic integrals,  the Liouville and the Lie
integrable systems as it happens in the classical case. The quantum
Liouville quadratic integrable systems are defined on a Liouville
manifold and the Schr\"odinger equation can be solved by separation
of variables in one coordinate system. The Lie integrable systems
are defined on a Lie manifold and are not generally separable ones
but the can be solved. Therefore there are superintegrable systems
with two quadratic integrals of motion not necessarily separable in
two coordinate systems. The quantum analogues of the two dimensional
superintegrable  systems with qua\-dra\-tic integrals of motion on a
manifold are classified by using the quadratic associative algebra
of the integrals of motion. There are six general fundamental
classes of quantum superintegrable systems corresponding to the
classical ones. Analytic formulas for the involved integrals are
 calculated in all the cases. All the known quantum
superintegrable systems are  classified as special cases of these six general
classes. The coefficients of the associative algebra of the general cases are
calculated. These coefficients are the same as the coefficients of the
classical case multiplied by $-\hbar^2$ plus
 quantum corrections of order $\hbar^4$ and $\hbar^6$.

\end{abstract}

\newpage

\section{Introduction}\label{sec:Introduction}

In classical mechanics, a superintegrable or completely integrable
is a Hamiltonian system with a maximum number of integrals. Two well
known examples are the harmonic oscillator and the Coulomb
potential. In the $N$-dimensi\-onal case the superintegrable system
has $2 N-1$ integrals, one among them is the Hamiltonian. In the two
dimensional case the number of integrals in a superintegrable system
is three.

The problem of complete classification of the superintegrable
systems is recently achieved  in the case where the integrals
of motion are quadratic equations of the momenta
 \cite{KalKrMiPo02,KaKrMi05_I,KaKrMi05_II}.
 All the known classical potentials can be classified in six classes \cite{DasYps06}.
  This paper is essentially the quantum version of the reference  \cite{DasYps06}.

In classical mechanics, if the potential is switched off, the problem is
reduced to find the manifolds which possess more than one integrals. This
problem was studied by Koenings in 19th century, see Darboux: \textit{Le\c{c}ons
sur la Th\'{e}orie G\'{e}n\'{e}rale des Surfaces} \cite{Darboux}. In the same
extremely detailed book Darboux classified the two dimensional manifolds with
geodesics accepting two quadratic integrals (the Hamiltonian and one integral
of motion)  in two main classes, the Liouville and the Lie manifolds\cite[vol
III, n$^{o}$ 596, p. 38]{Darboux}. In modern language the systems defined on a
manifold which are integrable with quadratic integrals of motion can be
classified in two classes the Liouville and the Lie systems
\cite{KaKrMi05_I,KaKrMi05_II,DasYps06}. The Koenigs' main result was that,
there are five classes of general forms of metrics,  whose the geodesics have
three integrals of motion (the Hamiltonian and two additional functionally
independent integrals). These metrics are called "formes essentielles" and they
depend on four parameters. All the metrics having more than two integrals of
motion, can be obtained as partial cases of these "formes essentielles" by
choosing appropriate values of the four parameters. The five classes of metrics
are tabulated in "Tableau VII" by Koenigs\cite[vol IV p.385]{Darboux}.

The study of  classical and quantum dimensional superintegrable systems with
quadratic integrals of motion started with the investigation of superintegrable
systems on the flat space and on spaces with constant curvature
\cite{Fris,Hietarinta87}. In many instances,  the notion of superintegrability
was associated to the notion of separability on two different coordinate
systems. All the separable classical and quantum systems are classified
\cite{KalKrMiPo02} and all the quadratic superintegrable systems in a flat
space, on a space with constant curvature and on spaces by revolution  are
investigated
\cite{KaMiPogo96,KaMiPogo00,KaMiPogo00b,KaKrPogo01,KaKrWin02,KaMiPogo02_PAN,KaKrMiWin03,Ranada97,RaSan99,RaSan02}.
If the Hamiltonian has   quadratic and  non trivial linear terms then the
superintegrability don't imply necessarily the separation of variables in two
different coordinate systems \cite{BeWin04}. This means that the notion of
superintegrability don't coincide with the notion of separabilty in two
coordinate systems.
 In this paper  the  existence of non separable two dimensional integrable systems
 with quadratic integrals of motion is discussed.

The superintegrable systems corresponding to the Koenigs' "formes essentielles"
were studied in \cite{KaKrMi05_I,KaKrMi05_II} where the integrals of motion are
calculated.  This "Tableau VII" of ref.  \cite{Darboux} was created by studying
the metrics possessing almost three integrals of motion on a manifold. In ref.
\cite{DasYps06} six classes of superintegrable systems are proved to exist.
This result is corroborated by classifying the Staeckel equivalent
superintegrable systems  \cite{Kress}.

Each special case of a classical superintegrable system can be generated by
fixing the associated constants  in the appropriate class of Poisson algebra.
This classification has been achieved by classifying all the possible Poisson
quadratic algebras of the integrals. In this paper we show that, to each
classical potential corresponds a quantum system and the Poisson algebra of
integrals is replaced by a quadratic associative algebra as it was studied in
\cite{BoDasKo93,BoDasKo94,Vinet,KaMiPogo96,Das01}. The coefficients of the
associative algebra are analytically calculated in all the cases as in ref.
\cite{DasYps06}.  These coefficients are the similar as those, which were
calculated in the classical case, but he are multiplied by $-\hbar^2$, the
quantum coefficients are differentiated from the classical case by the presence
of deformations of order $\hbar^4$ and $\hbar^6$, which represent the quantum
effects. The classification introduced in the classical case, which is based on
the properties of the associative algebra, is an alternative method to the
classifications based on the separation of variables \cite{KalKrMiPo02} or to
the study of the possible Darboux coefficients. These methods are very
efficient if the manifold is fixed
\cite{KaMiPogo96,KaMiPogo00,KaMiPogo00b,KaKrPogo01,KaKrWin02,KaMiPogo02_PAN,KaKrMiWin03,Ranada97,RaSan99,RaSan02}.
The situation is more complicated in the case of a general manifold
\cite{KaKrMi05_I,KaKrMi05_II,DasYps06}.

This paper is organized as follows: In section \ref{sec:QIS}  the well known
general forms of the quantum quadratic integrable systems on a manifold   are
derived for consistency reasons. The general solution of the Schr\"odinger
equation for the  Lie integrable systems is given.  In Section \ref{sec:QSS}
the quantum analogues of the classical superintegrable systems are discussed.
To every classical system corresponds a quantum system and the quantum systems
can be classified by classifying the corresponding quadratic associative
algebra of integrals. This classification is similar to the classification of
classic systems by using  the quadratic Poisson algebra of integrals. In
Section \ref{sec:Classification} the potential and the integrals of the six
classes of two dimensional quadratic superintegrable systems are calculated.
The coefficients of the associative quadratic algebra of integrals is
calculated and they are generated by the classical coefficients multiplied by
$-\hbar^2$ plus quantum corrections of order $\hbar^4$ and $\hbar^6$. Finally
in Section \ref{sec:Discussion} the results of this paper and the open problems
are shortly discussed.


\section{Quantum integrable systems} \label{sec:QIS}
Let us consider a two dimensional manifold with metric:
\begin{equation}\label{eq:uvmetric}
ds^2=E(u,v) du^2+2F(u,v) du dv+G(u,v) dv^2
\end{equation}
In this coordinate system the quantum Hamiltonian has the general form
\begin{equation}\label{eq:uvH}
H=-\frac{\hbar^2}{2}\triangle + V
\end{equation}
where $\triangle$ is the Laplacian (or Laplace Beltrami operator)
$$
  \begin{array}{rl}
\displaystyle \triangle= \textrm{div} \circ \;\textrm{grad}=&
\dfrac{1}{\sqrt{E G-F^2}} \; \partial_{u} \Bigl (\dfrac{G}{\sqrt{E
G-F^2}} \;
\partial_{u}-\dfrac{F}{\sqrt{E G-F^2}}\; \partial_{v} \Bigr )+ \\
 \displaystyle +&\dfrac{1}{\sqrt{E G-F^2}} \; \partial_{v} \Bigl (\dfrac{E}{\sqrt{E
G-F^2}} \;
\partial_{v}-\dfrac{F}{\sqrt{E G-F^2}}\; \partial_{u} \Bigr )
 \end{array}
$$
 The above metric
(\ref{eq:uvmetric}), using a conformal coordinate system, can be
written as:
\[
ds^2=g(x,y) \, dx dy
\]
In the conformal coordinate system the Hamiltonian (\ref{eq:uvH}) has the form
\begin{equation}\label{eq:xyH}
H=-\frac{\hbar^2}{2}\triangle + V=-\frac{\hbar^2
}{g(x,y)} \;
\partial_{xy}+V(x,y)
\end{equation}
If the system possesses a quadratic integral of motion $I$, then under the above mentioned
conformal coordinate system, the integral is an operator which has the form:
\[\label{eq:integral1}
\begin{split}
I=-A(x,y) \, \hbar^2 \,  \partial_{xx}- B(x,y) \, \hbar^2 \,
\partial_{yy}+2 \, \hbar^2  \displaystyle \, \frac{\beta(x,y)}{g(x,y)}\;
\partial_{xy}\\-\imag \hbar \, r(x,y)  \, \partial_{x} - \imag \hbar \, s(x,y)
\partial_{y}+Q(x,y)
\end{split}
\]
By definition the Lie bracket between $H$ and $I$ must vanish:
  \begin{equation}
 [ H,I ]=H I-I H=0 \label{eq:lie}
  \end{equation}
The above equation implies restrictions on the involved functions
of $H$ and $I$. So, the coefficients of the partial derivatives $\partial_{xxx}$ and
$\partial_{yyy}$ in equation (\ref{eq:lie}) must be zero:
 \begin{equation}\label{eq:AB}
 \begin{split}
\frac{\partial A(x,y)}{\partial y}=0 & \Rightarrow  A=A(x)\\
\frac{\partial B(x,y)}{\partial x}=0 & \Rightarrow  B=A(y)
 \end{split}
 \end{equation}
Similarly, for the coefficients of $\partial_{xx}$ and
$\partial_{yy}$ we have:
 \begin{equation}\label{eq:rs}
 \begin{split}
\frac{\partial r(x,y)}{\partial y}=0 & \Rightarrow  r=r(x)  \\
\frac{\partial s(x,y)}{\partial x}=0 & \Rightarrow  s=s(y)
 \end{split}
 \end{equation}
The vanishing of the coefficients of $\partial_{xxy}$ and
$\partial_{yyx}$ gives:
\begin{equation}\label{eq:beta1}
\frac{\partial \beta}{\partial y}=A(x)\frac{\partial g}{\partial
x}+\frac{g}{2}A'(x)
 \end{equation}
\begin{equation}\label{eq:beta2}
\frac{\partial \beta}{\partial x}=B(y)\frac{\partial g}{\partial
y}+\frac{g}{2}B'(y)
 \end{equation}
combining the last equations we get
\begin{equation}\label{eq:eqmetric}
g(x,y)(A''(x)-B''(y))-3 B'(y)\frac{\partial g}{\partial y}-2
B(y)\frac{\partial^2 g}{\partial y^2}+3 A'(x)\frac{\partial
g}{\partial x}+2 A(x)\frac{\partial^2 g}{\partial x^2}=0
\end{equation}
From the coefficient equation of $\partial_{xy}$ taking into
account the relations (\ref{eq:beta1})  and  (\ref{eq:beta2}), we
get
  \[
\begin{split}
 g(x,y)\Bigl ( -2\imag r'(x)-2 \imag s'(y)\, +
\hbar(A''(x)+B''(y)) \Bigr )+(-2 \imag s(y) + \hbar B'(y))
 \dfrac{\partial g}{\partial y}\,+ \\ +(-2\imag
r(x)\,  + \hbar A'(x))  \dfrac{\partial g}{\partial
x}=0
\end{split}
 \]
or
\begin{equation}\label{eq:linear_terms}
\frac{\partial}{\partial x} \biggl ( \Bigl ( -2 \imag r(x)+ \hbar
A'(x) \Bigr ) g(x,y) \biggr )+\frac{\partial}{\partial y} \biggl (
\Bigl ( -2 \imag s(y)+ \hbar B'(y) \Bigr ) g(x,y) \biggr )=0
\end{equation}
and if we choose
\begin{equation}\label{eq:choose_rs}
r(x)=-\frac{1}{2} \imag \hbar A'(x), \qquad s(y)=-\frac{1}{2}
\imag \hbar B'(y)
\end{equation}
the equation (\ref{eq:linear_terms}) is identically zero and the integral of
motion $I$ is written:
\begin{equation}\label{eq:I_final}
  \begin{split}
I=- \hbar^2 \, A(x)\, \partial_{xx}-\hbar^2 \, B(y)\,
\partial_{yy}+2 \hbar^2  \frac{\beta(x,y)}{g(x,y)}\;\partial_{xy} -\frac{\hbar^2}{2} \,A'(x)\, \partial_{x}- \\
- \frac{\hbar^2}{2} \,B'(y)\, \partial_{y}  +Q(x,y)
 \end{split}
\end{equation}

The coefficients of $\partial_{x}$ and $\partial_{y}$ in
(\ref{eq:lie}) must be zero, so:
\begin{eqnarray}
\frac{\partial Q}{\partial y}&=& 2 g(x,y) A(x)\frac{\partial
V}{\partial
x}-2\beta(x,y)\frac{\partial V}{\partial y} \label{eq:Q1}\\
\frac{\partial Q}{\partial x}&=&2 g(x,y) B(y)\frac{\partial
V}{\partial x}-2\beta(x,y)\frac{\partial V}{\partial y}
\label{eq:Q2}
\end{eqnarray}
The partial derivative over $x$ of equation (\ref{eq:Q1}) is equal to the derivative over $y$ of equation (\ref{eq:Q2}),  this equality   imply:
\begin{equation}
\label{eq:poten}\begin{split}
g(x,y)\Bigl ( 3B'(y) \frac{\partial V}{\partial y}+
2B(y)\frac{\partial^2 V}{\partial y^2}- 3A'(x)\frac{\partial
V}{\partial x}-2 A(x)\frac{\partial^2 V}{\partial x^2} \Bigr )+ \\
+ 4B(y)\frac{\partial g}{\partial y}\frac{\partial V }{\partial
y }-4A(x)\frac{\partial g}{\partial x}\frac{\partial V}{\partial
x}=0
\end{split}
\end{equation}
Equations (\ref{eq:eqmetric}) and (\ref{eq:poten}) are the same as in the
classical case \cite{DasYps06},  therefore we can distinguish two kinds of
quantum superintegrable systems. The class I or Liouville integrable systems
and the class II or Lie integrable systems.

\medskip

\noindent\textbf{Class I: or  Liouville systems:} $A(x)B(y)\neq 0$

\noindent We choose a new coordinate system $(\xi, \eta)$ with
$$\xi=\int \frac{dx}{\sqrt{A(x)}}, \qquad  \eta =\int \frac{dy}{\sqrt{B(y)}}$$
in this system the differential operators written as:
$$\partial_{x}=\frac{\partial \xi}{\partial x}\,
\partial_{\xi}=\frac{1}{\sqrt{A(x)}}\, \partial_{\xi}, \qquad \partial_{y}=\frac{\partial \eta}{\partial
y} \,\partial_{\eta}=\frac{1}{\sqrt{B(y)}}\, \partial_{\eta} $$
therefore the Hamiltonian $H$ and the integral $I$ have the next
form respectively:
$$
H=-\frac{\hbar^2 }{g(\xi,\eta)}\,\partial_{\xi \eta}+V(\xi,\eta)
$$
$$
I=-\hbar^2 \partial_{\xi \xi}-\hbar^2 \partial_{\eta \eta}+2
\hbar^2 \,\frac{\beta(\xi,\eta)}{g(\xi,\eta)}\, \partial_{\xi
\eta}+Q(\xi,\eta)
$$

The coordinates $(\xi,\eta)$ usually referred as {\emph{Liouville
coordinates}}, and the equation (\ref{eq:eqmetric}) is
considerably simplified: $$\frac{\partial^2 g}{\partial
\xi^2}-\frac{\partial^2 g}{\partial \eta^2}=0$$ with general
solution $$g(\xi,\eta)=F(\xi+\eta)+G(\xi-\eta)$$ Where $F(u),G(v)$
arbitrary functions.

Now, from equations (\ref{eq:beta1}), (\ref{eq:beta2}) we can
calculate the function $\beta(\xi,\eta)$. The above mentioned
equations are simplified and they written as:
$$\frac{\partial \beta}{\partial \xi}=\frac{\partial g}{\partial \eta}, \qquad
\frac{\partial \beta}{\partial \eta}=\frac{\partial g}{\partial \xi}$$
therefore eliminating $g(\xi,\eta)$ from the last equations we get
\[
\frac{\partial^2 \beta}{\partial \xi^2}-\frac{\partial^2 \beta}{\partial
\eta^2}=0\]
therefore
\[
\beta(\xi,\eta)=F(\xi+\eta)-G(\xi-\eta)
\]

The equation (\ref{eq:poten}) in Liouville coordinates is
$$\Bigl ( F(\xi+\eta)+G(\xi-\eta) \Bigr )(V_{\xi \xi}-V_{\eta
\eta})+2F'(\xi+\eta)(V_{\xi}-V_{\eta})+2G'(\xi-\eta)(V_{\xi}-V_{\eta})=0
$$
where with $V_{x},V_{y}$ etc. denoted the partial derivatives of
$V(\xi,\eta)$. The general solution of the last equation is $$
V(\xi,\eta)=\frac{f(\xi+\eta)+g(\xi-\eta)}{F(\xi+\eta)+G(\xi-\eta)}
$$
with $f(u),g(v)$ arbitrary functions. Final the function
$Q(\xi,\eta)$ easily calculated from equations (\ref{eq:Q1}),
(\ref{eq:Q2}):
$$Q(\xi,\eta)=4\frac{f(\xi+\eta)G(\xi-\eta)-g(\xi-\eta)F(\xi+\eta)}{F(\xi+\eta)+G(\xi-\eta)}
$$

Introducing the corresponding isothermic coordinate system $(u,v)$ which are defined by:
$$\xi=\frac{u+iv}{2}, \qquad \eta=\frac{u-iv}{2} $$
The Hamiltonian $H$ and the integral $I$ can be written as:
$$H=\frac{1}{F(u)+G(v)}\Bigl ( \left(f(u)-\hbar^2 \partial_{uu})+(g(v)-\hbar^2 \partial_{vv}\right)\Bigr )$$
 $$I=\frac{1}{F(u)+G(v)} \left(4 G(v)(f(u)-\hbar^2 \,\partial_{uu})-4 F(u)(g(v)-\hbar^2 \,\partial_{vv})\right)$$
The relations $$H\Psi=E\Psi, \qquad I\Psi=J\Psi $$ lead us to a
formula for the Schr\"odinger function $\Psi$:
$$
\frac{1}{F(u)+G(u)}(f(u)-\hbar^2
\partial_{uu})\Psi+\frac{1}{F(u)+G(u)}(g(v)-\hbar^2\partial_{vv})\Psi=E\Psi
$$
$$
\frac{4 G(v)}{F(u)+G(v)}(f(u)-\hbar^2
\partial_{uu})\Psi-\frac{4 F(u)}{F(u)+G(v)}(g(v)-\hbar^2 \partial_{vv})\Psi=J\Psi
$$
combining the last equations we get
$$4 (f(u)-\hbar^2 \, \partial_{uu})\Psi=(4F(u)E+J)\Psi $$ $$4 (g(v)-\hbar^2 \,\partial_{vv})\Psi=(4G(v)E-J)\Psi $$
and if we let $$\Psi(u,v)=U(u)V(v)$$ the above equations separate
their variables:
$$U''(u)=\frac{1}{4 \hbar^2}(4f(u)-4F(u)E-J)U(u) $$ $$V''(v)=\frac{1}{4 \hbar^2}(4g(v)-4G(v)E+J)V(v). $$
Therefore an integrable Class I (Liouville) quantum system can be solved by
separation of variables.

\bigskip

 \noindent\textbf{Class II or  Lie systems:} $B(y)=0$

\noindent Defining a new coordinate system
 \[
 \xi=\int \frac{dx}{\sqrt{A(x)}},
\qquad \eta=y \]
 the differential operators are
$$\partial_{x}=\frac{\partial \xi}{\partial x}\,
\partial_{\xi}=\frac{1}{\sqrt{A(x)}}\; \partial_{\xi}, \qquad \partial_{y}=\partial_{\eta}$$
In this case the Hamiltonian $H$ and the integral $I$ is written:
$$
H=-\hbar^2 \frac{1}{g(\xi,\eta)}\,
\partial_{\xi\eta}+V(\xi,\eta)
$$
$$
I=-\hbar^2 \, \partial_{\xi \xi}+2\,\hbar^2 \frac{\beta
(\xi,\eta)}{g(\xi,\eta)}\,
\partial_{\xi \eta}+Q(\xi,\eta)
$$
This specific coordinate system $(\xi,\eta)$ is called  {\emph{Lie coordinate
system}}. In Lie coordinates equation (\ref{eq:eqmetric}) is written:
$$\frac{\partial^2 g}{\partial \xi^2}=0 $$
The general solution of the last equation is $$g(\xi,\eta)=F(\eta)
\xi+G(\eta)
$$
This metric defines a class of manifolds, which are called Lie manifolds.
Similarly to the previous case the equations (\ref{eq:beta1}),
(\ref{eq:beta2}) lead us to the next system:
$$
\begin{array}{l}
\displaystyle \frac{\partial \beta}{\partial \eta}=\frac{\partial
g}{\partial \xi} \\ \displaystyle
 \frac{\partial \beta}{\partial \xi}=0
\end{array}
$$
Taking into account the above expression for $g(\xi,\eta)$, we have
$$\beta(\xi,\eta)=\int\limits_{\eta_0}^{\eta} F(\eta')\,d\eta' $$
Equation (\ref{eq:poten}) is written:
 $$\Bigl (F(\eta) \xi+G(\eta) \Bigr ) \frac{\partial^2V}{\partial \xi^2}+2F(\eta)\frac{\partial V}{\partial \xi}=0 $$
with general solution given by the next expression
$$V(\xi,\eta)=\frac{f(\eta)\,\xi+g(\eta)}{F(\eta)\,\xi+G(\eta)}
$$
where $F(\eta),\, G(\eta),\, f(\eta),\, g(\eta)$ arbitrary functions.
Equations (\ref{eq:Q1}),  (\ref{eq:Q2}) imply:
$$Q(\xi,\eta)=\dfrac{-2(f(\eta)\xi +g(\eta))}{F(\eta)\,\xi+G(\eta)}
\int\limits_{\eta_0}^{\eta} F(\eta')\,
d\eta'+2\int\limits_{\eta_0}^{\eta} f(\eta') \, d\eta'
$$
The relations $$H\Psi=E\Psi, \qquad I\Psi=J\Psi $$ lead us to the Schr\"odinger
equation:
\begin{equation}\label{eq:LieSE}
-\frac{\hbar^2 \Psi_{\xi
\eta}}{F(\eta) \xi +G(\eta)}+\frac{f(\eta) \xi +g(\eta)}{F(\eta)
\xi +G(\eta)} \Psi= E \Psi
\end{equation}
The solution $\Psi$ of this equation satisfies the following equation:
\[
-\hbar^2 \Psi_{\xi \xi}+2\frac{\hbar^2 \int F(\eta) d\eta}{F(\eta) \xi +G(\eta)} \Psi_{\xi \eta}-2 \frac{f(\eta) \xi +g(\eta)}{F(\eta) \xi +G(\eta)}
\int F(\eta) d\eta \Psi +2 \Psi \int f(\eta) d\eta =J \Psi
\]
Taking into consideration Schr\"odinger equation (\ref{eq:LieSE}), the above
equation can be written:
\begin{equation}\label{eq:LieIS}
-\hbar^2 \Psi_{\xi \xi}= \left( J  +  2 \left( E
\int\limits_{\eta_0}^{\eta} F(\eta') d\eta'-
\int\limits_{\eta_0}^{\eta} f(\eta') d\eta' \right)\right)\Psi
\end{equation}
Let put
\[
\Pi(\eta)=  J  +  2 \left( E \int\limits_{\eta_0}^{\eta} F(\eta')
d\eta'- \int\limits_{\eta_0}^{\eta} f(\eta') d\eta' \right),\quad
p(\eta)=\sqrt{\left| \Pi(\eta)\right|}
\]
We can find solutions comparable to the solutions given in WKB
problems.
\begin{equation}\label{eq:sol1}
\begin{split}
\mbox{If  }&\;  \Pi(\eta)\ge 0 \;\rightsquigarrow  \Psi(\eta)=
A(\eta)\,  {\rm e}^{  \imag\, \xi \,p(\eta)/\hbar}  +B(\eta)
{\rm e}^{ -\imag\,\xi\,
p(\eta)/\hbar}\\
\mbox{where }&\quad A(\eta)=\exp\left[ -\int\limits_{\eta_0}^{\eta}
\frac{- \hbar\,f(\eta' )   + E\,\hbar\,F(\eta' ) -
    \imag \,\left(  - g(\eta' ) + E\,G(\eta' ) \right) \,
     {p(\eta')}}{\hbar\,
    \Pi(\eta') } \, d\eta'\right]\\
\mbox{and}&\quad B(\eta)=\exp\left[-\int\limits_{\eta_0}^{\eta}
\frac{- \hbar\,f(\eta' )   + E\,\hbar\,F(\eta' ) +
    \imag \,\left(  - g(\eta' ) + E\,G(\eta' ) \right) \,
     {p(\eta')}}{\hbar\,
    \Pi(\eta') }  \, d\eta'\right]\\
\mbox{If  }&\;  \Pi(\eta)< 0 \;\rightsquigarrow  \Psi(\eta)=
a(\eta)\,  {\rm e}^{   \xi \,p(\eta)/\hbar}  + b(\eta)
{\rm e}^{ -\xi\,
p(\eta)/\hbar}\\
\mbox{where }&\quad  a(\eta)=\exp\left[- \int\limits_{\eta_0}^{\eta}
\frac{ -\hbar\,f(\eta' )   + E\,\hbar\,F(\eta' ) -
    \left( - g(\eta' ) + E\,G(\eta' ) \right) \,
     {p(\eta')}}{\hbar\,
    \Pi(\eta') } \, d\eta'\right]\\
\mbox{and }&\quad b(\eta)=\exp\left[- \int\limits_{\eta_0}^{\eta}
\frac{- \hbar\,f(\eta' )   + E\,\hbar\,F(\eta' ) +
    \left(-g(\eta' ) + E\,G(\eta' ) \right) \,
     {p(\eta')}}{\hbar\,
    \Pi(\eta') } \, d\eta'\right]
\end{split}
\end{equation}

 If we
compare equations  (\ref{eq:eqmetric}) and (\ref{eq:poten}) with the
corresponding ones for the classical two dimensional systems with quadratic
integrals, they are indeed the same. Therefore we have shown the following
Proposition:
\begin{prop}\label{prop:Classical_Quantum}
Any two dimensional classical integrable system on a manifold with quadratic
integrals corresponds  to a quantum integrable system. There are two classes of
quantum integrable systems, the Liouville systems defined on a Liouville
manifold (Class I systems). These integrable systems can be solved by
separation of variables . The other class of integrable systems are the Lie
ones, which are defined on a Lie manifold (Class II systems). These integrable
systems cannot solved generally by separation of variables but they can be
solved by using WKB like solutions.
\end{prop}

\section{Quantum superintegrable systems} \label{sec:QSS}

A system is superintegrable on a two dimentional manifold if it has three
functionally independent integrals of motion $H,A$ and $B$. Let us consider a
Hamiltonian $H$ which in Liouville coordinate system has the form:
$$H=\frac{-\hbar^2}{g(\xi, \eta)}\;\partial_{\xi \eta}+V(\xi,\eta)$$ From the
previous section we know that the integral of motion $A$ of an integrable
system, can be written in two specific forms analogous to two distinct
coordinate systems which are: $$
\begin{array}{l}
 \displaystyle A=-\hbar^2 \partial_{\xi \xi}-\hbar^2 \partial_{\eta
\eta}+2\hbar^2\,\frac{\beta(\xi,\eta)}{g(\xi,\eta)}\;
\partial_{\xi \eta}+Q(\xi,\eta) \qquad \text{(Liouville system)}\\
\displaystyle A=-\hbar^2 \partial_{\xi \xi}+2\hbar^2\;
\frac{\beta(\xi ,\eta)}{g(\xi ,\eta)}\; \partial_{\xi \eta}+ Q(\xi
,\eta)\qquad \qquad \qquad \text{(Lie system)}
\end{array}
$$
Let us consider the second integral of motion, in Liouville
coordinates,
 of the form (\ref{eq:I_final}) namely,
$$B=-\hbar^2 A(\xi)\partial_{\xi \xi}-\hbar^2 B(\eta)
\partial_{\eta \eta}+2\hbar^2 \frac{\beta(\xi ,\eta)}{g(\xi
,\eta)}\; \partial_{\xi \eta}- \frac{\hbar^2}{2} A'(\xi)
\partial_{\xi}- \frac{\hbar^2}{2} B'(\eta) \partial_{\eta}+Q(\xi,\eta)
$$ therefore the following relations are satisfied:
$$
[H,A]=[H,B]=0
$$
In this paper we study the quantum superintegrable systems where the integrals $A,B$ and $C$ satisfy the quadratic associative
algebra as in ref. \cite{Das01}:
\begin{equation}\label{eq:AssocAlg}
 \begin{array}{l}
\left [A,B \right ] = C  \\
\left[ A,C \right ]=\alpha A^2+\beta B^2+ \gamma \{ A,B \}+\delta A+ \epsilon B +\zeta \\
 \left [B,C \right ] = a A^2-\gamma B^2-\alpha  \{ A, B\}  +d
A-\delta B+z
 \end{array}
 \end{equation}
where $a, \gamma, \alpha, \beta $ are constants and
$$
\begin{array}{c}
 \delta=\delta(H)=\delta_{1} H+\delta_{0}, \\
 \epsilon=\epsilon_{1} H+\epsilon_{0}, \\
 \zeta=\zeta(H)=\zeta_{2} H^2+\zeta_{1} H+\zeta_{0}, \\
 d=d(H)=d_{1} H+d_{0}, \\
 z=z(H)=z_{2} H^2+z_{1} H+z_{0,}
\end{array}
$$
with $\delta_{i},\epsilon_{i},\zeta_{i},d_{i} $ and $z_{i}$
constants. The Casimir of the above quadratic algebra is given by
the expression:
\begin{equation} \label{eq:Casimir}
\begin{split}
  K=C^2-\alpha \{ A^2,B\}-\gamma\{ A,B^2 \}+\left(\alpha
\gamma -\delta+\frac{a \beta}{3}\right)\{A,B\}
-\\
-\frac{2 \beta}{3} B^3 +\left(\gamma^2-\epsilon-\dfrac{\alpha
\beta}{3}\right) B^2 +\left(-\gamma
\delta+2\zeta-\frac{\beta  d}{3}\right)B  +\\
+\frac{2 a}{3} A^3+\left(d+\frac{a
\gamma}{3}+\alpha^2\right) A^2+\left(\frac{a \epsilon}{3}+\alpha \delta+2 z\right)
A  \end{split}
\end{equation}
Since $B$ is an integral of motion there exist a new coordinate system $(X,Y)$,
let it be the Liouville one, in which the integral $B$ has the following
 form:
 $$B=-\hbar^2 \partial_{XX}-\hbar^2 \partial_{YY}+2 \hbar^2 \frac{\widetilde{\beta}(X,Y)}{\widetilde{g}(X,Y)}\; \partial_{XY}+\widetilde{Q}(X,Y) $$
where $\widetilde{\beta}(X,Y), \widetilde{Q}(X,Y)$ the functions
$\beta(\xi,\eta), Q(\xi,\eta)$ in $(X,Y)$ coordinates and,
$$\widetilde{g}(X,Y)=\frac{g(\xi,\eta)}{\sqrt{A(\xi)}\sqrt{B(\eta)}}$$

The integral $B$ can be replaced by a linear combination of the integrals $A,\,B$ and $H$ and the coefficient $\beta$ can be put always to be $0$.
In this case,    the coefficients of
$(\partial_{\xi})^6$ and $(\partial_{\eta})^6$ must vanish in the Casimir (\ref{eq:Casimir}), so the
following relations are true:
\begin{equation}\label{eq:Casimir6derivatives}
 \begin{split}
6\hbar^2 (A'(\xi))^2=a-3\gamma A^2(\xi)+3\alpha A(\xi)  \\
6\hbar^2 (B'(\eta))^2=a-3\gamma B^2(\xi)+3\alpha B(\xi)
\end{split}
\end{equation}

The superintegrable systems on a manifold can be classified by the
solutions of the last equations. These equations differ from those
of classical case only in a multiplier $-\hbar^2$ as a coefficient
of $A'(\xi)$ and $B'(\eta)$.  There are two classes of superintegrable systems.
The class I systems corresponding to the case where both the integrals $A$ and $B$ are Liouville integrals and in the class II systems where the first integral $A$ is of Lie type and the second is a Liouville integral. Regarding equations (\ref{eq:Casimir6derivatives})  the superintegrable systems can be classified in the following six subclasses:

\begin{table}[h]\label{tab:Classes}

\[
\begin{array}{|c|c|c|c|c|c|}
\hline
 & \alpha & \gamma &a &  A(\xi) & B(\eta)\\
\hline\hline
I_1    & 0  & 0& 6\hbar^2 & \xi & \eta\\
\hline
I_2    &   -8\hbar^2  & 0 & 0& \xi^2 & \eta^2\\
\hline
I_3    & 32 \hbar^2  & -8\hbar^2& 0 & (e^{\xi}+e^{-\xi})^2 & (e^{\eta}+e^{-\eta})^2 \\
\hline
II_1    & 0 & 0 & 0 & 1 & 0\\
\hline
II_2    &  0 & 0& 6\hbar^2 & \xi & 0\\
\hline
II_3    & -8\hbar^2  & 0 & 0 & \xi^2& 0\\
\hline\hline
\end{array}
\]

\caption{\textsf{Classification Table of quantum quadratic superintegrable systems ($\beta=0$)}}
\end{table}

 The associative algebra of integrals is characterized by the coefficients of the quantum
superintegrable systems,  these coefficients  are exactly the same to the
classical Poisson algebra constants multiplied to $-\hbar^2$.

Let us begin by the known solutions $A(\xi),\; B(\eta)$ of equations
(\ref{eq:Casimir6derivatives}).
 Taking into consideration that:
\[
ds^2=g(\xi,\, \eta) \,d\xi d\eta= \left(F(\xi+\eta)+G(\xi-\eta)\right)   \,d\xi
d\eta
\]
 Equation (\ref{eq:eqmetric}) is written for the Class I superintegrable systems
\begin{equation}\label{eq:Diff_betaI}
\begin{array}{l} \left( A''(\xi)-B''(\eta) \right) \left( F(\xi+\eta)+ G(\xi-\eta) \right)
+\\
+ 3 A'(\xi) \left( F'(\xi+\eta)+ G'(\xi-\eta) \right) -3 B'(\eta) \left( F'(\xi+\eta)-
G'(\xi-\eta) \right)+\\+2  \left(A(\xi)-B(\eta)\right) \left( F''(\xi+\eta)+
G''(\xi-\eta) \right)=0
\end{array}
\end{equation}
From the above equation the functions $F(u)$ and $G(v)$ are calculated by
separation of variables for each subclass of quantum superintegrable system. We
should notice that this equation is the same as in the classical case. Taking
into consideration the general form of the potential in the case of Liouville
integrable systems
\[
V(\xi,\eta)= \dfrac{f(\xi+\eta)+g(\xi-\eta)}{F(\xi+\eta)+G(\xi-\eta)}
\]
equation (\ref{eq:poten}) leads to a differential equation for
  the functions $f(u)$ and $g(v)$ which are involved in the
  definition   of the potential:
\begin{equation}\label{eq:complicated_eq}
\begin{array}{l}
f(\xi + \eta)\,\{ -3\,B'(\eta)\,
      \left( F'(\xi + \eta) - G'(\xi - \eta) \right)  +
     3\,A'(\xi)\,\left( F'(\xi + \eta) + G'(\xi - \eta) \right)  +\\
     +
     2\,\left( A(\xi) - B(\eta) \right) \,
      \left( F''(\xi + \eta) + G''(\xi - \eta) \right) \}  +\\
 + g(\xi - \eta)\,\{ -3\,B'(\eta)\,
      \left( F'(\xi + \eta) - G'(\xi - \eta) \right)  +
     3\,A'(\xi)\,\left( F'(\xi + \eta) + G'(\xi - \eta) \right)  +\\
    + 2\,\left( A(\xi) - B(\eta) \right) \,
      \left( F''(\xi + \eta) + G''(\xi - \eta) \right) \}  -\\
      -
  \left( F(\xi + \eta) + G(\xi - \eta) \right) \,
  \{ -3\,B'(\eta)\,
      \left( f'(\xi + \eta) - g'(\xi - \eta) \right)  +\\
     +3\,A'(\xi)\,\left( f'(\xi + \eta) +
        g'(\xi - \eta) \right)  +
     2\,\left( A(\xi) - B(\eta) \right) \,
      \left( f''(\xi + \eta) + g''(\xi - \eta) \right)
     \}=0
\end{array}
\end{equation}
We can eliminate from the above equation the functions $F(u)$ and $G(v)$, which
satisfy equation (\ref{eq:Diff_betaI}) and finally the functions involved in
the definition of the potential satisfy the following equation:
\begin{equation}\label{eq:DiffQI}
\begin{array}{l} \left( A''(\xi)-B''(\eta) \right) \left( f(\xi+\eta)+ g(\xi-\eta) \right)
+\\
+ 3 A'(\xi) \left( g'(\xi+\eta)+ g'(\xi-\eta) \right) -3 B'(\eta) \left( g'(\xi+\eta)-
g'(\xi-\eta) \right)+\\+2  \left(A(\xi)-B(\eta)\right) \left( f''(\xi+\eta)+
g''(\xi-\eta) \right)=0
\end{array}
\end{equation}
This equation is indeed the same as (\ref{eq:Diff_betaI}).  The
general solutions of equations  (\ref{eq:Diff_betaI})  and  (\ref{eq:DiffQI}) are the same as in the classical case see ref \cite{DasYps06}. These solutions are studied in Section \ref{sec:Classification}.

For the class II integrable systems
\[
ds^2= \left(F(\eta) \xi +G(\eta) \right) d\xi d\eta
\]
 Equation (\ref{eq:eqmetric}) is written:
 \begin{equation}\label{eq:Diff_betaII}
\begin{array}{l} \left( A''(\xi)-B''(\eta) \right) \left( F(\eta)\xi + G(\eta) \right)
+\\
+ 3 A'(\xi) F(\eta) -3 B'(\eta) \left( F'(\eta) \xi+ G'(\eta) \right)+2
\left(A(\xi)-B(\eta)\right) \left( F''(\eta)  \xi+ G''(\eta) \right)=0
\end{array}
\end{equation}
From the above equation the functions $F(\xi)$ and $G(\eta)$ are calculated.
This equation is the same as in the classical case. Taking
into consideration the general form of the potential in the case of Lie
integrable systems
\[
V(\xi,\eta)= \dfrac{f(\eta)\xi+g(\eta)}{F(\eta)\xi+G(\eta)}
\]
equation (\ref{eq:poten}) leads to a complicated differential equation for
  the functions $f(\xi)$ and $g(\eta)$ analogous to
  equation (\ref{eq:complicated_eq}) and after some algebra we find:
\begin{equation}\label{eq:DiffQII}
\begin{array}{l} \left( A''(\xi)-B''(\eta) \right) \left( f(\eta)\xi + g(\eta) \right)
+\\
+ 3 A'(\xi) f(\eta) -3 B'(\eta) \left( f'(\eta) \xi+ g'(\eta) \right)+2
\left(A(\xi)-B(\eta)\right) \left( f''(\eta)  \xi+ g''(\eta) \right)=0
\end{array}
\end{equation}
The similarity of the pair of equations  (\ref{eq:Diff_betaI}), (\ref{eq:DiffQI}) and (\ref{eq:Diff_betaII}), (\ref{eq:DiffQII}) to the corresponding ones in the classical case \cite{DasYps06} leads to the following Proposition:
\\
\begin{prop}\label{prop:QSuper}
Each classical  superintegrable system with quadratic integrals  corresponds to a quantum superintegrable system. There are six independent classes of quantum superintegrable integrals. The potentials, integrals and the coefficients of the associative algebra of the integrals (\ref{eq:AssocAlg}) can be analytically calculated.
\end{prop}

\section{Classification of two dimensional superintegrable
systems with two quadratic integrals of
motion}\label{sec:Classification}

In this section we give the analytical solutions for the different classes of
superintegrable systems. As it was shown in Proposition \ref{prop:QSuper},
there are two general classes of superintegrable systems, each class has three
subclasses. In this section the solutions of equations (\ref{eq:Diff_betaI})
and (\ref{eq:DiffQI})  and the coefficients of the associative integral algebra
are calculated.

\subsection{Class I superintegrable systems}

\subsubsection{Subclass I${}_1$ of superintegrable systems }
\[
A(\xi)=\xi, \qquad B(\eta)=\eta
\]
\begin{equation}\label{eq:FGfgVa1}
\begin{array}{ll}
 \displaystyle F(u)=4 \lambda \,u^2\,+ \,\kappa \,u
\,+\,{\nu}/{2}, &
\displaystyle G(v)=-\lambda \,v^2\, + \,{\mu}/{ v^2 }\,+\,{\nu}/{2}\\

 \displaystyle f(u)=4 \ell \,u^2 \,+\, k\, u
+{n}/{2}, & \displaystyle g(v)=-\ell\, v^2\, + \,{m}/{ v^2}\,
+\,{n}/{2}\\
\end{array}
\end{equation}

\[
\begin{array}{ll}
ds^2=g(\xi,\eta)\, d\xi\, d\eta, &
g(\xi,\eta)= F(\xi+\eta)+G(\xi-\eta)\\
\\
\displaystyle H=\frac{-\hbar^2}{g(\xi,\eta)} \; \partial_{\xi
\eta}+V(\xi,\eta) &\displaystyle
V(\xi,\eta)=\frac{w(\xi,\eta)}{g(\xi,\eta)}, \quad
w(\xi,\eta)=f(\xi+\eta)+g(\xi-\eta)
\end{array}
\]
The other integral of motion is:
\[
\begin{array}{rl}
\displaystyle A=-\hbar^2 \partial_{\xi \xi}-\hbar^2 \partial_{\eta \eta} + &\displaystyle 2 \hbar^2 \frac{F(\xi+\eta)-G(\xi-\eta)}{F(\xi+\eta)+G(\xi-\eta)} \; \partial_{\xi \eta}+\\
+ & \displaystyle
4\frac{f(\xi+\eta)\,G(\xi-\eta)\,-\,g(\xi-\eta)\,
F(\xi+\eta)}{F(\xi+\eta)\,+\,G(\xi-\eta)}
\end{array}
\]
We introduce the functions:
\begin{equation}\label{eq:FGfgtildeVa1}
\begin{array}{l}\displaystyle \widetilde{F}(u)=\frac{\,\ \lambda u^6
}{256}+\frac{\,\ \kappa u^4 }{128} + \frac{\,\ \nu u^2}{16}
 -\frac{\mu }{u^2}
 \\
\\
  \displaystyle
\widetilde{G}(v)=-
  \frac{\,\ \lambda v^6}{256}-\frac{\,\ \kappa v^4}{128}  -
  \frac{\,\ \nu v^2}{16}+ \frac{\mu }{v^2}
\\
\\
\displaystyle \widetilde{f}(u)=\frac{\,\ \ell\,u^6}{256}+
\frac{\,\ k\,u^4}{128}+
\frac{\,\ n\,u^2}{16}-\frac{m}{u^2} \\
\\
     \displaystyle
\widetilde{g}(v)=- \frac{\,\ \ell\,v^6}{256} -
  \frac{\,\ k\,v^4}{128} -
\frac{\,\ n\,v^2}{16}+\frac{m}{v^2}
\end{array}
\end{equation}
The second integral of motion is:
\[
\begin{array}{rl}
\displaystyle  B=& \displaystyle  - \hbar^2 \partial_{X X} -
\hbar^2
\partial_{Y Y} + 2 \hbar^2 \;
\frac{\widetilde{F}(X+Y)-\widetilde{G}(X-Y)}{\widetilde{F}(X+Y)+\widetilde{G}(X-Y)}\; \partial_{X Y}+\\
&\displaystyle   + 4 \frac{\widetilde{f}(X+Y) \widetilde{G}(X-Y)-
\widetilde{g}(X-Y)\widetilde{F}(X+Y)}{\widetilde{F}(X+Y)+\widetilde{G}(X-Y)}
\end{array}
\]
where
\[
X=2 \sqrt{\xi}, \quad \partial_{X}= \sqrt{\xi}\; \partial_{\xi},
\quad Y=2 \sqrt{\eta}, \quad \partial_{Y}=\sqrt{ \eta} \;
\partial_{\eta}
\]
The constants of the Poisson algebra are:
\[
\begin{array}{l}
\alpha=0 \quad\beta=0 \quad \gamma=0 \quad \delta=-16 \hbar^2
 (\kappa H -k) \quad
 \epsilon=-256 \hbar^2 (  \lambda  H-\ell )\\
\zeta=32 \hbar^2 (\kappa H -k)  (\nu  H -n) \quad
a=6 \hbar^2 \qquad d=-8 ( \nu H-n) \\
z=-8 \hbar^2 ( \nu  H -n)^2+  128 \hbar^2 ( \mu H - m) (\lambda H
-\ell)-96 \hbar^4(\lambda H -\ell)
\end{array}
\]
\begin{eqnarray}
K&=&-32 \hbar^2
 (\nu  H-n)^3  - 512 \hbar^2 (\lambda H -\ell)( \nu H -n)( \mu H -m)  +  \nonumber\\
 &&   +64 \hbar^2 (\kappa H  -k)^2 (\mu H -m)-640 \hbar^4 (\lambda H -\ell)( \nu H -n)+48 \hbar^4 (\kappa H  -k)^2 \nonumber
\end{eqnarray}

\subsubsection{Subclass I${}_2$ of superintegrable systems }
\[
A(\xi)=\xi^2, \qquad B(\eta)=\eta^2
\]
\begin{equation}\label{eq:FGfgVa2}
\begin{array}{ll}
\displaystyle F(u)=
  \lambda\,u^2\, +\frac{\kappa }{u^2}  + \frac{\nu }{2}, &
\displaystyle G(v)= - \lambda \,v^2\,   +
  \frac{\mu }{v^2} + \frac{\nu }{2}\\
\\
\displaystyle f(u)=
  \ell\,u^2\, +\frac{k }{u^2}  + \frac{n }{2}, &
\displaystyle g(v)= - \ell\,v^2\,  +
  \frac{m }{v^2} + \frac{n }{2}
\end{array}
\end{equation}
\[
\begin{array}{ll}
ds^2=g(\xi,\eta)\, d\xi\, d\eta, &
g(\xi,\eta)= F(\xi+\eta)+G(\xi-\eta)\\
\\
\displaystyle H=\frac{-\hbar^2}{g(\xi,\eta)}\;
\partial_{\xi \eta}+V(\xi,\eta) &\displaystyle
V(\xi,\eta)=\frac{w(\xi,\eta)}{g(\xi,\eta)}, \quad
w(\xi,\eta)=f(\xi+\eta)+g(\xi-\eta)
\end{array}
\]
The other integral of motion is:
\[
\begin{array}{rl}
\displaystyle A=-\hbar^2\partial_{\xi \xi}-\hbar^2\partial_{\eta \eta} + &\displaystyle 2 \hbar^2 \frac{F(\xi+\eta)-G(\xi-\eta)}{F(\xi+\eta)+G(\xi-\eta)}\; \partial_{\xi \eta} +\\
+& \displaystyle 4\frac{f(\xi+\eta)\,G(\xi-\eta)\,-\,g(\xi-\eta)\,
F(\xi+\eta)}{F(\xi+\eta)\,+\,G(\xi-\eta)}
\end{array}
\]
We introduce the functions:
\begin{equation}\label{eq:FGfgtildeVa2}
\begin{array}{l}
\displaystyle \widetilde{F}(u)=4\,\lambda\,e^{2\,u}+\nu\,e^u ,
\quad
  \displaystyle
\widetilde{G}(v)=\frac{\kappa\,\ e^v}
   {{\left( 1 + e^v \right) }^2} +
  \frac{\mu\,\ e^v }
   {{\left( -1 + e^v \right) }^2}
\\
\\
\displaystyle \widetilde{f}(u)=4\,\ell\, e^{2\,u} +n\, e^u , \quad
     \displaystyle
\widetilde{g}(v)=\frac{k\, e^v} {{\left(1 + e^v \right)}^2} +
  \frac{m\, e^v}
{{\left(-1 + e^v \right)}^2}
\end{array}
\end{equation}
The second integral of motion is:
\[
\begin{array}{rl}
\displaystyle  B=& \displaystyle
-\hbar^2\partial_{XX}-\hbar^2\partial_{YY} + 2 \hbar^2 \,
\frac{\widetilde{F}(X+Y)-\widetilde{G}(X-Y)}{\widetilde{F}(X+Y)+\widetilde{G}(X-Y)}\; \partial_{XY}+\\
&\displaystyle   + 4 \frac{\widetilde{f}(X+Y) \widetilde{G}(X-Y)-
\widetilde{g}(X-Y)\widetilde{F}(X+Y)}{\widetilde{F}(X+Y)+\widetilde{G}(X-Y)}
\end{array}
\]
where
\[
X= \ln\xi,   \quad \partial_{X}= \xi\,\partial_{\xi}, \quad
Y=\ln\eta, \quad \partial_{Y}=\eta\, \partial_{\eta}
\]
The constants of the Poisson algebra are:
\[
\begin{array}{l}
\alpha=-8 \hbar^2\quad\beta=0 \qquad \gamma=0\quad \delta=0\quad
 \epsilon=-256 \hbar^2 (   \lambda  H-\ell)\\
\zeta=32 \hbar^2 (\nu H -n)^2-256 \hbar^2 ( \lambda H - \ell)
\left(  (\mu H -m)  - (\kappa H -k) \right)+128 \hbar^4 (\lambda
H-\ell)
\\
a=0 \qquad d=16 \hbar^4 \qquad z=-32 \hbar^2 \,\left( (\kappa H  -
k )  + (\mu H-m) \right) \,
   (\nu \,H-n)
\end{array}
\]

\[
\begin{array}{l}
K= -256  \hbar^2 \,(\lambda \, H -\ell)
    \left( (\kappa H  - k)  + (\mu H -m)  \right) ^2 -
    128  \hbar^2 \left(( \kappa H - k )
     -(\mu H-m)  \right) \, \\
      {(\nu  H -n) }^2
      + 128 \hbar^4 ((\nu H-n)^2+4(H\lambda -\ell) ((\kappa H-k)-(\mu H-m) )
      )+4 \hbar^6 (\lambda H-\ell)
\end{array}
\]

\subsubsection{Subclass I${}_3$ of superintegrable systems }
\[
A(\xi)=(e^{\xi}+e^{-\xi})^2, \qquad B(\eta)=(e^{\eta}+e^{-\eta})^2
\]
\begin{equation}\label{eq:FGfgVa3}
\begin{array}{ll}
 \displaystyle F(u)=\frac{\kappa\,e^{2\,u} }{{\left( -1 + e^{2\,u} \right) }^2} +
  \frac{\,\lambda\,e^u\,\left( 1 + e^{2\,u} \right) }
   {{\left( -1 + e^{2\,u} \right) }^2}, &
\displaystyle G(v)=\frac{\mu\,e^{2\,v} }{{\left( -1 + e^{2\,v}
\right) }^2} +
  \frac{\,\nu\,e^v\,\left( 1 + e^{2\,v} \right) }
   {{\left( -1 + e^{2\,v} \right) }^2}\\  \\

 \displaystyle f(u)=\frac{k\,e^{2\,u} }{{\left( -1 + e^{2\,u} \right) }^2} +
  \frac{\,\ell\,e^u\,\left( 1 + e^{2\,u} \right) }
   {{\left( -1 + e^{2\,u} \right) }^2}, & \displaystyle g(v)=\frac{m\,e^{2\,v} }{{\left( -1 + e^{2\,v}
\right) }^2} +
  \frac{\,n\,e^v\,\left( 1 + e^{2\,v} \right) }
   {{\left( -1 + e^{2\,v} \right) }^2}\\
\end{array}
\end{equation}
\[
\begin{array}{ll}
ds^2=g(\xi,\eta)\, d\xi\, d\eta, &
g(\xi,\eta)= F(\xi+\eta)+G(\xi-\eta)\\
\\
\displaystyle H= \frac{-\hbar^2}{g(\xi,\eta)}\,
\partial_{\xi \eta}+V(\xi,\eta) &\displaystyle
V(\xi,\eta)=\frac{w(\xi,\eta)}{g(\xi,\eta)}, \quad
w(\xi,\eta)=f(\xi+\eta)+g(\xi-\eta)
\end{array}
\]
The other integral of motion is:
\[
\begin{array}{rl}
\displaystyle A=-\hbar^2 \partial_{\xi \xi}-\hbar^2 \partial_{\eta \eta} + &\displaystyle 2 \hbar^2 \frac{F(\xi+\eta)-G(\xi-\eta)}{F(\xi+\eta)+G(\xi-\eta)}\; \partial_{\xi \eta} +\\
+& \displaystyle 4\frac{f(\xi+\eta)\,G(\xi-\eta)\,-\,g(\xi-\eta)\,
F(\xi+\eta)}{F(\xi+\eta)\,+\,G(\xi-\eta)}
\end{array}
\]
We introduce the functions:
\begin{equation}\label{eq:FGfgtildeVa3}
\begin{array}{l}
\displaystyle \widetilde{F}(u)=\frac{\left( \kappa  + 2\,\lambda
\right)}{4}\,\tan^2  u  +
  \frac{2\,\nu\,-\mu }{4}\,\cot^2 u + \frac{\lambda  + \nu }{2}
 \\
\\
  \displaystyle
\widetilde{G}(v)=\frac{\left(2\,\lambda -\kappa   \right) }{4}
\,\tan^2 v + \frac{\mu  + 2\,\nu }{4}\,\cot^2  v+
  \frac{\lambda  + \nu }{2}\\
\\
\displaystyle \widetilde{f}(u)=\frac{\left(k+ 2\,\ell \right)
}{4}\,\tan^2  u +
  \frac{2\,n\,- m }{4}\,\cot^2u + \frac{\ell + n }{2}\\
\\
     \displaystyle
\widetilde{g}(v)=\frac{\left(2\,\ell - k \right) }{4}\,\tan^2 v +
\frac{m + 2\,n }{4}\,\cot^2 v+
  \frac{\ell + n }{2}
\end{array}
\end{equation}
The second integral of motion is:
\[
\begin{array}{rl}
\displaystyle  B=& \displaystyle  -\hbar^2 \partial_{XX}-\hbar^2
\partial_{YY} + 2 \hbar^2 \,
\frac{\widetilde{F}(X+Y)-\widetilde{G}(X-Y)}{\widetilde{F}(X+Y)+\widetilde{G}(X-Y)}\; \partial_{XY}+\\
&\displaystyle   + 4 \frac{\widetilde{f}(X+Y) \widetilde{G}(X-Y)-
\widetilde{g}(X-Y)\widetilde{F}(X+Y)}{\widetilde{F}(X+Y)+\widetilde{G}(X-Y)}
\end{array}
\]

where
$$
X=\arctan (e^{\xi}) ,\quad \partial_X= (e^{\xi}+e^{-\xi})  \,
\partial_\xi, \quad    Y=\arctan (e^{\eta}),\quad \partial_Y=
(e^{\eta}+e^{-\eta})\, \partial_\eta
$$

The constants of the Poisson algebra are:
\[
\begin{array}{l}
\alpha=32 \hbar^2 \quad\beta=0 \quad \gamma=-8 \hbar^2 \quad
\delta=32 \hbar^4 \quad
 \epsilon=-16 \hbar^4 \quad
\zeta=32 \hbar^2\, \left( \lambda H - \ell \right) \left( \nu H
-n\right)
\\
a=0 \qquad d=-64 \hbar^2\,\left( \kappa H - k \right)+64
\hbar^2\,\left( \mu H -
    \mu  \right) +256 \hbar^4\\
z=-32 \hbar^2 \, {\left( (\lambda  -
        \nu) H - (\ell-n)   \right) }^2 +32 \hbar^2 \, \left( (\kappa   H -k\right)  \,
       \left(\mu H - m  \right ) )+32\hbar^4(\mu H-m)-\\ \qquad -32 \hbar^4(\kappa H-k)
\end{array}
\]
\[
\begin{array}{l}
K=-64 \hbar^2 \,\left( \kappa H -k\right) \,{\left(\nu   H - n
\right)}^2 +64 \hbar^2{\left( \lambda H -\ell\right) }^
        2\,\left(\mu H -m \right)-512 \hbar^4 (\nu H-n)\\ \qquad (\lambda H-l)-64 \hbar^4(\mu H-m)(\kappa H-k)
        +128 \hbar^4 (\lambda H-l)^2+128 \hbar^4 (\nu H-n)+\\ \qquad +128 \hbar^6 (\kappa H-k)-128 \hbar^6 (\mu H-m)
\end{array}
\]

\subsection{Class II superintegrable systems}
\subsubsection{Subclass II$_{1}$ of superintegrable systems }
\[
A(\xi)=1, \qquad B(\eta)=1
\]
\begin{equation}\label{eq:FGfgVb1}
\begin{array}{ll}
 \displaystyle F(\eta)=\kappa\,\eta + \lambda, &
\displaystyle G(\eta)=\mu\, \eta  + \nu\\
 \displaystyle f(\eta)=k\,\eta + \ell, & \displaystyle g(\eta)=m\, \eta + n\\
\end{array}
\end{equation}
\[
\begin{array}{ll}
ds^2=g(\xi,\eta)\, d\xi\, d\eta, &
g(\xi,\eta)=\xi\, F(\eta)+G(\eta)\\
\\
\displaystyle H=\frac{- \hbar^2}{g(\xi,\eta)}\;
\partial_{\xi \eta}+V(\xi,\eta) &\displaystyle
V(\xi,\eta)=\frac{w(\xi,\eta)}{g(\xi,\eta)}, \quad
w(\xi,\eta)=\xi\, f(\eta)+g(\eta)
\end{array}
\]
The other integral of motion is:
\[
\begin{array}{rl}
\displaystyle A = - \hbar^2 \partial_{\xi \xi} + \frac{2 \hbar^2
\,
     \int\limits_{\eta_0}^\eta F(\eta' )\,d\eta' }{g(\xi ,\eta )}\;\partial_{\xi \eta}  -
  \frac{2\,\left( \xi \,f(\eta )+ g(\eta )
       \right) \,\int\limits_{\eta_0}^\eta F(\eta' )\,d\eta' }{g(\xi ,
     \eta )} +  2\,\int\limits_{\eta_0}^\eta f(\eta' )\,d\eta'
\end{array}
\]
We introduce the functions:
\begin{equation}\label{eq:FGfgtildeVb1}
\begin{array}{l}
\displaystyle \widetilde{F}(u)=\frac{\kappa \,u^2}{4} +
  \frac{\left( \lambda  + \mu  \right)\,u }{2} +
  \frac{\nu }{2}
 \\
\\
  \displaystyle
\widetilde{G}(v)=-\frac{\kappa\, v^2 }{4} + \frac{\left( \lambda -
\mu \right) \,v }{2} +
  \frac{\nu }{2}\\
\\
\displaystyle \widetilde{f}(u)=\frac{k\,u^2}{4} +
  \frac{\left( \ell + m  \right)\,u }{2} +
  \frac{n}{2}\\
\\
     \displaystyle
\widetilde{g}(v)=-\frac{k\, v^2 }{4} + \frac{\left( \ell - m
\right) \,v }{2} +
  \frac{n }{2}
\end{array}
\end{equation}
The second integral of motion is:
\[
\begin{array}{rl}
\displaystyle B=&\displaystyle - \hbar^2 \partial_{\xi \xi} -
\hbar^2 \partial_{\eta \eta}+
 2  \hbar^2\,  \frac{
     \widetilde{F}(\xi + \eta ) -
       \widetilde{G}( \xi - \eta  ) }{\widetilde{F}(\xi + \eta ) + \widetilde{G}( \xi -\eta )}\,\partial_{\xi \eta} +\\  \\
& \displaystyle +\, 4\, \frac{  \widetilde{f}(\xi+ \eta
)\,\widetilde{G}(\xi -\eta )-\widetilde{g}( \xi -\eta
)\,\widetilde{F}(\xi + \eta )
  }
     {\widetilde{F}(\xi + \eta ) + \widetilde{G}(\xi -\eta )}
\end{array}
\]

The constants of the Poisson algebra are:
\[
\begin{array}{l}
\alpha=0 \quad\beta=0 \quad \gamma=0 \quad \delta=-8
\hbar^2(k-\kappa\, H) \quad
 \epsilon=0\quad
\zeta=-8  \hbar^2 \,\left( \lambda H - \ell  \right)^2
\\
a=0 \qquad d=-16  \hbar^2 (k-\kappa\, H )\quad z=-8  \hbar^2
\left( \lambda H - \ell \right)^2+8  \hbar^2 \left( \mu H - m
\right)^2
\end{array}
\]
\[
K=-16 \hbar^2 \, \left( \nu H - n  \right)^2 \left( \kappa H - k
\right) +32 \hbar^2 \, \left( \lambda H - \ell  \right) \left( \mu
H - m \right)\left( \nu H - n  \right)-16 \hbar^4 (k-\kappa H)^2
\]

\subsubsection{Subclass II$_{2}$ of superintegrable systems }
\[
A(\xi)=\xi, \qquad B(\eta)=\eta
\]
\begin{equation}\label{eq:FGfgVb2}
\begin{array}{ll}
 \displaystyle F(\eta)=\frac{\kappa }{{\sqrt{\eta }}} + \lambda, &
\displaystyle G(\eta)=3\,\kappa\,{\sqrt{\eta }}  + \lambda\,\eta +
\frac{\mu }{{\sqrt{\eta }}} + \nu \\ \\
 \displaystyle f(\eta)=\frac{k }{{\sqrt{\eta }}} + \ell, &
 \displaystyle g(\eta)=3\,k\,{\sqrt{\eta }}  + \ell\,\eta
+  \frac{m }{{\sqrt{\eta }}} + n
\end{array}
\end{equation}
\[
\begin{array}{ll}
ds^2=g(\xi,\eta)\, d\xi\, d\eta, &
g(\xi,\eta)=\xi\, F(\eta)+G(\eta)\\
\\
\displaystyle H=\frac{-\hbar^2}{g(\xi,\eta)}\;
\partial_{\xi \eta}+V(\xi,\eta) &\displaystyle
V(\xi,\eta)=\frac{w(\xi,\eta)}{g(\xi,\eta)}, \quad
w(\xi,\eta)=\xi\, f(\eta)+g(\eta)
\end{array}
\]
The other integral of motion is:
\[
\begin{array}{rl}
\displaystyle A = - \hbar^2 \partial_{\xi \xi}+
  \frac{2 \hbar^2
     \int\limits_{\eta_0}^\eta F(\eta' )\,d\eta' }{g(\xi ,\eta )}\; \partial_{\xi \eta}  -
  \frac{2\,\left( \xi \,f(\eta )+ g(\eta )
       \right) \,\int\limits_{\eta_0}^\eta F(\eta' )\,d\eta' }{g(\xi ,
     \eta )} +  2\,\int\limits_{\eta_0}^\eta f(\eta' )\,d\eta'
\end{array}
\]
We introduce the functions:
\begin{equation}\label{eq:FGfgtildeVb2}
\begin{array}{l}
\displaystyle \widetilde{F}(u)= \frac{\,\lambda\, u^4 }{128}
+\frac{\,\kappa\,u^3 }{16} +\frac{\,\nu\, u^2 }{16}+
  \frac{\,\mu\, u\ }{4}

 \\
\\
  \displaystyle
\widetilde{G}(v)=-
  \frac{\,\lambda\,v^4 }{128}+\frac{\,\kappa\,v^3 }{16}  +
  \frac{\,\mu\,v }{4} -
  \frac{\,\nu\,v^2 }{16}\\
\\
\displaystyle \widetilde{f}(u)=\frac{\,\ell\, u^4 }{128}
+\frac{\,k\,u^3 }{16} +\frac{\,n\, u^2 }{16}+
  \frac{\,m\, u\ }{4}
\\   \\
     \displaystyle
\widetilde{g}(v)=-
  \frac{\,\ell\,v^4 }{128}+\frac{\,k\,v^3 }{16}  +
  \frac{\,m\,v }{4} -
  \frac{\,n\,v^2 }{16}
\end{array}
\end{equation}
The second integral of motion is:
\[
\begin{array}{rl}
\displaystyle  B=& \displaystyle  - \hbar^2\partial_{XX}-\hbar^2
\partial_{YY} + 2 \hbar^2
\frac{\widetilde{F}(X+Y)-\widetilde{G}(X-Y)}{\widetilde{F}(X+Y)+\widetilde{G}(X-Y)}\; \partial_{XY}+\\
&\displaystyle   + 4 \frac{\widetilde{f}(X+Y) \widetilde{G}(X-Y)-
\widetilde{g}(X-Y)\widetilde{F}(X+Y)}{\widetilde{F}(X+Y)+\widetilde{G}(X-Y)}
\end{array}
\]
where
\[
X=2 \sqrt{\xi}, \quad \partial_{X}= \sqrt{\xi} \;\partial_{\xi},
\quad Y=2 \sqrt{\eta}, \quad \partial_{Y}=\sqrt{ \eta}\;
\partial_{\eta}
\]

The constants of the Poisson algebra are:
\[
\begin{array}{l}
\alpha=0 \quad\beta=0 \quad \gamma=0 \quad \delta=-4 \hbar^2
\,(\ell - \lambda\, H) \qquad
 \epsilon=0\quad
\zeta=-8 \hbar^2 {\left(\kappa  H - k \right)}^2
\\
a= 6 \hbar^2 \qquad d= -8 \hbar^2 \,(\nu\, H -n )\quad z=  8
\hbar^2 \, \left(\kappa\,H-k\right) \left( \mu H -m \right)  +2
\hbar^2 {\, \left(\nu  H -n  \right)}^2
\end{array}
\]
\[
K=-8 \hbar^2 \,\left( \lambda\,H - \ell \right)  \left( \mu H -m
\right) ^2+ 16 \hbar^2 \, \left(\kappa\,H - k)  (\mu\ H -m\right)
\left( \nu\, H -n\right)-4 \hbar^4 (\ell -H\lambda)^2
\]

\subsubsection{Subclass II$_{3}$ of superintegrable systems }
\[
A(\xi)={\xi}^2, \qquad B(\eta)={\eta}^2
\]
\begin{equation}\label{eq:FGfgVb3}
\begin{array}{ll}
 \displaystyle F(\eta)=\lambda\, \eta + \frac{\kappa }{{\eta }^3} , &
\displaystyle G(\eta)=\nu + \frac{\mu }{{\eta }^2}  \\  \\

 \displaystyle f(\eta)=\ell\, \eta + \frac{k}{{\eta }^3}, &
 \displaystyle g(\eta)=n + \frac{m}{{\eta }^2}
\end{array}
\end{equation}
\[
\begin{array}{ll}
ds^2=g(\xi,\eta)\, d\xi\, d\eta, &
g(\xi,\eta)=\xi\, F(\eta)+G(\eta)\\
\\
\displaystyle H=\frac{-\hbar^2 }{g(\xi,\eta)}\;
\partial_{\xi \eta}+V(\xi,\eta) &\displaystyle
V(\xi,\eta)=\frac{w(\xi,\eta)}{g(\xi,\eta)}, \quad
w(\xi,\eta)=\xi\, f(\eta)+g(\eta)
\end{array}
\]
The other integral of motion is:
\[
\begin{array}{rl}
\displaystyle A = - \hbar^2 \partial_{\xi \xi} +
  \frac{2 \hbar^2 \,
     \int\limits_{\eta_0}^\eta F(\eta' )\,d\eta' }{g(\xi ,\eta )} \; \partial_{\xi \eta}  -
  \frac{2\,\left( \xi \,f(\eta )+ g(\eta )
       \right) \,\int\limits_{\eta_0}^\eta F(\eta' )\,d\eta' }{g(\xi ,
     \eta )} +  2\,\int\limits_{\eta_0}^\eta f(\eta' )\,d\eta'
\end{array}
\]
We introduce the functions:
\begin{equation}\label{eq:FGfgtildeVb3}
\begin{array}{l}
\displaystyle \widetilde{F}(u)=\lambda\,e^{2\,u}  + \nu\, e^u,
\quad
  \displaystyle
\widetilde{G}(v)=\kappa\, e^{2\,v}  +\mu\, e^v\\
\displaystyle \widetilde{f}(u)=\ell\,e^{2\,u}  + n\, e^u, \quad
     \displaystyle
\widetilde{g}(v)=k\, e^{2\,v}  +m\, e^v
\end{array}
\end{equation}
The second integral of motion is:
\[
\begin{array}{rl}
\displaystyle  B=& \displaystyle  - \hbar^2
\partial_{XX}-\hbar^2  \partial_{YY} + 2 \hbar^2 \,
\frac{\widetilde{F}(X+Y)-\widetilde{G}(X-Y)}{\widetilde{F}(X+Y)+\widetilde{G}(X-Y)}\; \partial_{XY}+\\
&\displaystyle   + 4 \frac{\widetilde{f}(X+Y) \widetilde{G}(X-Y)-
\widetilde{g}(X-Y)\widetilde{F}(X+Y)}{\widetilde{F}(X+Y)+\widetilde{G}(X-Y)}
\end{array}
\]
where
\[
X= \ln \xi,   \quad \partial_{X}= \xi\,\partial_{\xi}, \quad Y=\ln
\eta, \quad \partial_{Y}=\eta\, \partial_{\eta}
\]
The constants of the Poisson algebra are:
\[
\begin{array}{l}
\alpha=-8 \hbar^2 , \quad\beta=0, \quad \gamma=0, \quad \delta=0,
\quad
 \epsilon=0,\quad
\zeta=-32 \hbar^2 \, \left ( \kappa\,H -k \right)
\left(\lambda\,H-\ell\right ),
\\
a=0, \quad d=16 \hbar^4 , \quad z=-32 \hbar^2  \, \left(
\mu\,H-m\right)\left( \nu\, H-n \right)
\end{array}
\]
\[
K=-64 \hbar^2 \, \left(\lambda\, H-\ell\right){\left(\mu  H
-m\right)}^2 +64 \hbar^2 \left( \kappa\, H -k\right) {\left(\nu  H
-n \right)}^2+64 \hbar^4 (k-H\kappa)(H \lambda-\ell)
\]

\section{Discussion}\label{sec:Discussion}

In this paper the quantum superintegrable  system with quadratic integrals of
motion are classified  in six classes as the classical ones. The potentials,
integrals and the coefficients of the associative algebra of the integrals
(\ref{eq:AssocAlg}) are calculated.  The problem of finding and classify of the
quadratic superintegrable systems is related to the algebraic problem of
classifying quadratic Poisson algebras in the classical case. In the quantum
case the notion of superintegrability is ambigious, therefore the quantum
analogues of the classical systems are constructed and the quantum analogues of
the Poisson quadratic algebra is the associative quadratic algebra of
integrals. The superintegrability don't imply necessarily separation of
variables in two different systems because the superintegrable systems
possessing a Liouville and a Lie integral are not necessarily separable in two
systems, but they are solvable as it was proved in Section \ref{sec:QIS}. The
existence of nonseparable superintegrable systems has been shown in the case of
superintegrable systems with Hamiltonian having nontrivial linear in momenta
terms plus the quadratic ones \cite{BeWin04}.

There are several open problems arising from the study of the two dimensional
superintegrable systems.  We shall indicate some of these problems

\begin{enumerate}
\item The calculation of the energy eigenvalues of these systems
 is  related to the finite dimensional representations of the quadratic algebra of integrals
   \cite{Das01}. Therefore the mathematical problem of the study of  representation theory of
   quadratic and polynomial extensions of the Lie algebras is an open mathematical problem. The representations of the unitary representations of the quadratic algebra related to the four classical superintegrable systems have been studied in  \cite{Das01}, using the techniques of lowering and uppering operators which is related to the deformed oscillator algebra \cite{Das}. The importance of quadratic or cubic assosiative algebras in the integrable and solvable problems and the discussion of a  representation theory can be found in \cite{GLZ-1992,Zhedanov93}.

\item
The classification of two dimensional systems with a cubic  and a quadratic
integral is an open problem. The cubic problems have been investigated recently
\cite{GravelWin02,Gravel04,KaKrMi05_I,KaKrMi05_II,KaKrMi05_III,KaKrMi06_IV,Tsyg00_TMP,Tsyg00_JPA}.

\item
Recently the problem of the classification of the classical and quantum three dimensional in conformally flat spaces has been investigated \cite{KaKrMi05_III} and \cite{KaKrMi06_IV}. The general theory of the superintegrability in a general manifold is one of the most interesting open problems.

\end{enumerate}

\bigskip

\noindent ACKNOWLEDGMENTS:
 This article is part of a project supported by PYTHAGORAS II/80897

\end{document}